\documentclass{optica-article}

\journal{opticajournal} 

\articletype{Research Article}

\usepackage{lineno}

\newcommand{\fig}[1]{Fig. \ref{#1}}

\newcommand{\eq}[1]{eq. (\ref{#1})}
\newcommand{\Eq}[1]{Eq. (\ref{#1})}

\begin{document}

\title{Filter shaping in integrated photonics using non-adiabatic control}

\author{Abdullah Al-Abiad\authormark{1,*}, and Pierre Colman\authormark{1}}

\address{\authormark{1} Universit\'{e} Bourgogne Europe, CNRS, Laboratoire Interdisciplinaire Carnot de Bourgogne, UMR6303, Dijon (FRANCE) }

\email{\authormark{*}abdullah.f.alabiad@gmail.com , pierre.colman@u-bourgogne.fr} 


\begin{abstract*}
We demonstrate that an array of integrated coupled waveguides can be designed to simultaneously perform the spectral filtering and routing of light. The framework that we present here operates in a regime opposite to the shortcut-to-adiabaticity techniques, but we show it nevertheless also only requires smooth, slow, and minute modulations of the array's parameters. Our design strategy rely on the analogy which exists between the temporal evolution of an Hamiltonian and the propagation of light in an array of coupled integrated waveguides, mimicking spatially in photonics phenomena analogue to the electronic transitions. Interestingly, the chromatic dispersion in photonics restores the correspondence between spatial propagation effect and wavelength dependence. Therefore our framework is able to fully reproduces temporal effects such as resonant transitions, making it suitable for filtering applications. As a design involving modulations, it shares similarities with Bragg-Gratings (BGs). But it adds some extra features that make it complementary in terms of applications and functionality. First it enables routing with the seamless management of multi-waveguide configuration. And secondly the framework we present requires only wavelength-scale modulations, order of magnitudes slower than for BGs. As a result, the effects we present here may not interfere with BGs’ geometry and functionality; and could even be combined with the latter.
\end{abstract*}

\section{Introduction}
 Spectral filters are key building blocks in integrated photonics, with applications in numerous areas including telecommunications \cite{fernandez2024nanophotonic,zhang2025bragg,little1997microring}, sensing \cite{liang2005highly,Kannojia2025}, microwave photonics \cite{Burla2013, liu2022tunable, Sun2022, Nasehi2025}, among others. Typical design strategies for filters rely either on optical interferences \cite{fernandez2024nanophotonic,halir2011high}, or on resonances \cite{little1997microring,liu2025compact,deng2017tunable}. The Multimodal Interferometer (MMI) is perhaps the most well-known design illustrating the former strategy \cite{halir2011high}. In this system, light is fed into a waveguide supporting several optical modes, whose mutual interference creates at the output a complex pattern that changes rapidly with wavelength. The energy can be collected by means of several ports that then constitute separate channels with distinctive frequency responses. Careful design would allow one to obtain complementary inputs and outputs, thus achieving a wavelength demultiplexer (WDM). Instead of relying on the interference between several co-propagating modes, interference can also be created between the forward- and backward-propagating waves in the so-called Bragg gratings geometry (WBGs, \cite{bruckerhoff2025general}). Compared to MMIs, WBGs are very flexible and more straightforward in their design, allowing the design of filters with variable wavelength selectivity \cite{strain2014multi}. Compactness is also a key feature. In this respect, Bragg filters can also be cascaded, or even superimposed, to result in fully tunable multi-wavelength filters. The design of Bragg filters obeys a few quite straightforward rules, making them quite easy to optimize and control \cite{bruckerhoff2025general,Ctyroky2018,Skaar2001}. 
 \par Although this technology is already used widely, a few design challenges still remain open. Firstly, the fundamental physics of WBGs implies the presence of reflected waves. Therefore the latter must be somehow managed, because either they contain the useful signal, or upstream elements could also be negatively impacted by an optical return originating from the WBGs. If we consider that the use of circulators is still a very arduous task in integrated photonics, this issue requires the use of extra elements. This involve in peculiar the use of MMI or couplers, which have their own constraints in terms of bandwidth and selectivity  \cite{fernandez2024nanophotonic,zhang2025bragg}. This renders the overall design actually very complex, contrasting greatly with its underlying more simple working principle. Secondly, if the principles governing the design of WBGs are well understood \cite{Praena2024,bruckerhoff2025general}, the design of complex filters becomes rapidly a quite difficult task. Indeed the interference of several Bragg gratings operating over the about same wavelength range and at the same position is usually not considered : a discrete by-block design strategy is preferred, which result in sub-optimal and less compact systems. Note that because BGS requires sub-wavelength modulation, the collocation of several filters at the same place would create some aliasing effects, and in turn requires nm-scale geometrical features, which are very difficult to fabricate \cite{Paul2017}.
 
 \par In a different approach, the propagation of light in an array of coupled waveguide obeys an equation analogue to the time evolution of an Hamiltonian. This analogy between optical propagation and quantum mechanics has been at the heart of various demonstrations and quantum-inspired photonic phenomena \cite{Longhi_2009,Szameit2009, Biagioni2008, DellaValle2007, Longhi2006a, Longhi2005 }.  We propose in this article a design technique relying on the optical analogue of electronic transitions, where light can be selectively scattered between defined photonic states. This framework was previously proposed as a mean to manage the pathway of light inside compact array of coupled waveguide \cite{Sheveleva2022}, but that study only considered continuous wave propagation and overlooked the spectral filtering capability offered by such an architecture. Our present work focuses therefore on the spectral selectivity offered by these photonic state transitions. We demonstrate that photonic transitions can be combined to form complex spectral shaping functions; while also providing concomitant spatial separation, hence routing. 
   
 \par We first describe the concept underlying the \textit{non-adiabatic control}, namely the control of transition of light between photonic states. We will also introduce the few key parameters that permit the inverse design of the photonic structure (geometry) based on the objectives, which are defined in terms of optical features. In a second part, we present several typical examples of spectral shaping that can be obtained, focusing on the manner in which different features can be combined and merged. In a third part, we illustrate more specifically the light routing capability. Finally we present  more advanced features and we discuss how our current design could be further improved and optimized.
 
\section{Non-adiabatic control}
 The set of equations governing the evolution of light in an array of coupled integrated waveguides has the same structure as the Schrödinger equation describing the temporal evolution of a quantum system \cite{Longhi_2009}. We consider here the optical analogue of electronic transitions. To understand how such an equivalent effect could be controlled in photonics, we consider first in \eq{eq1} the general equation governing the propagation of light in a varying (modulated) array:
 \begin{equation}
 	\frac{d\psi(z)}{dz}
 	=
 	i
 	\begin{bmatrix}
 		K_1(z) & \kappa_{12}(z) & 0 \\
 		\kappa_{12}(z) & K_2(z) & \ddots \\
 		0 & \ddots & \ddots
 	\end{bmatrix}
 	\psi(z)
 	=
 	i M(z)\psi(z)
 	\label{eq1}
 \end{equation}
 With $\psi(z)=\begin{bmatrix}
 	E_1(z)\\
 	\vdots\\
 	E_n(z)
 \end{bmatrix} $; $|E_i(z)|^2$ represents the power flowing in the $i^{th}$ waveguide, $\kappa_{ij}$ is the coupling constant between the $i^{th}$ and the $j^{th}$ waveguides. $K_{i}$ is the propagation constant of $i^{th}$ waveguide if taken isolated. The $\beta_{i=[1,n]}$ are the propagation wavevectors of  the eigenmodes associated with a static array; and they are obtained by the diagonalization of $M(z)=V(z)B(z)V^\dagger(z)$
 with $B(z)=
 \begin{bmatrix}
 \beta_1(z) &    0    & 0 \\
 0          & \ddots  &  0 \\
 0          &  0      & \beta_n(z)
 \end{bmatrix}$
 
 We define $\psi(z)=V(z)\Psi(z)$, where $\Psi(z)$ encodes the distribution of light among the photonic supermodes (eigen-solutions on a snapshot basis). Projected on this new basis, \eq{eq1} becomes:
\begin{equation}
 \frac{d\Psi(z)}{dz}
 =
 \imath B(z)\Psi(z)
 -
 \imath V^\dagger(z)\frac{dV(z)}{dz}\Psi(z)
 	\label{eq2}
\end{equation}
 
\Eq{eq2} reveals explicitly the impact of longitudinal variations through the appearance of the $V^\dagger d_z V$ operator. Note that because \eq{eq2} is expressed in the instantaneous eigen-mode basis -also called the snapshot basis- of the Hamiltonian, off-diagonal terms correspond to the scattering of light  from one photonic mode to another. In the frame of the so-called shortcut-to-adiabaticity theory \cite{Taras2021,Ho2015,Tseng2014,Paspalakis2006}, this operator must be kept with a minimal impact; while the objective here would be to maximize its overall effect. By opposition to the shortcut-to-adiabaticity, we call the this regime the \textit{Non-adiabatic} regime.

\subsection{Inverse Design}
\par The operator involves a spatial derivative. As a consequence very rapid variations of the array's parameters would result in various scattering in-between photonic modes. However this would happen in an uncontrolled manner. Interestingly the $V^\dagger d_z V$ operator can be inverse designed, namely its value can be first fixed and the corresponding array evolution $M(z)$ can then be computed. This requires solving the following equation:
 \begin{equation}
 	\frac{\partial M(z)}{\partial z}
 	=
 	V(z)\,\frac{\partial B(z)}{\partial z}\,V^\dagger(z)
 	+
 	M(z)\,V(z)\left[V^\dagger \frac{dV}{dz}\right]_c V^\dagger(z)
 	-
 	V(z)\left[V^\dagger\frac{dV}{dz}\right]_c V^\dagger(z)\,M(z)
 	\label{eq3}
 \end{equation}
 
\begin{figure}[h]
 	\centering
 	\includegraphics[width=\linewidth]{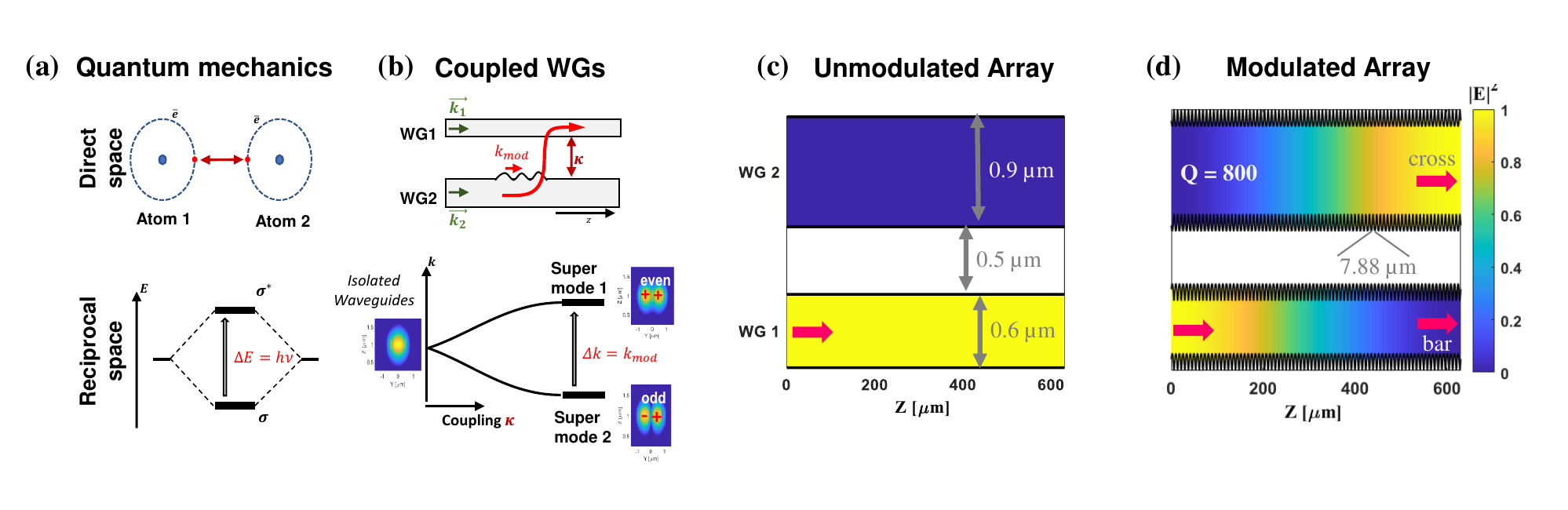}
 	\caption{(a) Effect of atom coupling on electronic levels. (b) Schematic equivalence for a coupled waveguide array, where spatial (propagation) coordinates replace temporal effects. (c) Propagation of light in a dissymmetric array. Array parameters, as used throughout this article, are indicated. (d) Propagation in the same array as in (c), but modulated with a period of 7.88 $\mu$m, hence enabling the transfer of light between the waveguides. Red arrows indicate the position of light input and outputs.}
 	\label{fig1}
 \end{figure}

 Once the parameters evolution is known, the actual geometry can be inferred using a parameter-to-geometry look-up table. A typical example is shown in \fig{fig1}(c) where light can be selectively scattered into another waveguide – namely another photonic mode – only under the action of the modulation.  Albeit the evolution is so-called non-adiabatic because it involves the transfer of energy between different photonic states, it can be triggered by extremely slow – but cumulative– modulations. Here the modulation period is about $8~{\mu}m$
 
 The geometrical variations of the structures corresponding to the solution of \eq{eq3} are also shown. Solutions of \eq{eq3} are periodic modulations, whose period  $\Lambda_{ij,\lambda}=2\pi/|\beta_i-\beta_j |$ compensates for the phase-mismatch between the photonic modes. This is expected from physics, and the proposed architecture shares then a strong similarity with WBGs. However, in contrast to the latter where the modulation is imposed on the waveguide itself, the modulation in an array of waveguides can involve more degrees of freedom, including both the waveguides' width or the inter-waveguide spacing, each parameters being modulated with its own prescribed amplitude en phase. A typical illustration is shown in \fig{fig3}-(c). Indeed the photonic modes not only have their own propagation constant, but also their own symmetry selection rules that the modulation must satisfy. As a result  over the range of situations that we tested numerically, we found that even if the geometry does not exactly match the prescribed parameters evolution defined by \eq{eq3}, the flow of light still follows closely the desired evolution. This robustness comes from the fact that small errors in the designs usually do not respects the correct periodicity, not symmetry; and therefore their actual impact is severely reduced. 
 
 \subsection{Design rules}
 \par More sophisticated routing solutions can also be obtained \cite{Sheveleva2022}. Note that \eq{eq3} can be controlled by simple objective parameters, therefore greatly facilitating the design process. Indeed the coupling length can be expressed as $L_{ij,\lambda}=\frac{Q\,\Lambda_{ij,\lambda}}{2}$, where Q characterizes the quality factor of the transition, which can be freely set. Q can be also seen as the ratio of the coupling length over the modulation period. Interestingly, Q is also directly related to the spectral bandwidth of the transition. Indeed if the propagation of light along the array is equivalent to the time evolution of an Hamiltonian, the large chromatic dispersion naturally present in integrated photonics introduces a strong wavelength dependence for the design. As a result, the system shows at its output a spectral dependence very similar as for the time evolution of an Hamiltonian; and hence enables spectral filtering and other temporal-like effects to take place. The maximal relative coupling efficiency $\eta_{\max}$ reads as:
 \begin{equation}
 	\eta_{\max}
 	=
 	\frac{1}{1+\left(\frac{(\Lambda_{ij,\lambda}-\Lambda_{0}) Q_0}{2\Lambda_{ij,\lambda}}\right)^2}
 	\label{eq4}
 \end{equation}
 Where $\Lambda_{ij,\lambda}$ is the modulation required at the considered wavelength $\lambda$ in order to couple the \textit{i} and \textit{j} states, and $\Lambda_{0}$ is the modulation imprinted on the array. Eventually $\Lambda_{0}=\Lambda_{ij,\lambda_0}$ is set so that the array works optimally at the frequency $\lambda_0$. If the modulation is then imposed over a length $L_{mod}$, the relative amount of energy transferred amounts then to $T = \eta_{\max}\,\sin^2\!\left(\frac{\pi}{2}\frac{L_{\mathrm{mod}}}{L_{ij,\lambda}}\right)$.
 These rules of thumb allow quite simple prototyping of the filter, whose corresponding exact modulation features are first obtained by solving \eq{eq3}. Note that the expression for $\Lambda_{ij,\lambda}$, and especially its wavelength's dependency can be quite complex. 
 
 \section{Tunable filters}

\par We will now illustrate typical spectral features that can be obtained using this deterministic transfer of light between optical modes, a.k.a. \textit{Non-adiabatic control}. This will be the occasion to showcase how different operations can be superposed spatially and operate either independently on different spectral ranges, or combined to result in custom filters with tunable shape.

\subsection{Numerical Model}
\par All the results presented in this article are numerical simulations based on the effective parameters propagation equation \eq{eq1}. Indeed the overall dimensions of the structures considered here renders any direct full 3D simulation impractical. That said, the effective parameters are directly derived from 3D simulations done using the MPB package \cite{MPB}, which serves as a reference look-up table. This means that the output of the inverse design equation \eq{eq3} is not directly fed to \eq{eq1}; but that the closest corresponding geometry is first defined; and the latter serves then to estimate the effective parameters used in \eq{eq1}. For geometrical array's configurations which were not exactly computed in the look-up table, effective parameters are then interpolated using our latest coupled-waveguide model \cite{MPB}. The results are therefore intended to reflect realistic geometry. To be more specific, we consider 350nm-height waveguides made in Silicon Nitride ($n_{SiN}\approx1.8$). The waveguide are embedded in a silicon matrix ($n_{SiO2}=1.45$). Waveguides have widths ranging from 550nm to 900nm, corresponding to effective indices of about $n_{eff}\approx1.65$. For sake of simplicity, we consider about constant array, meaning we impose $\partial_zB(z)=0$ when solving \eq{eq3}. More general situations, and their implications, are further discussed in the last section of the paper.

\subsection{Spectral Filters}
\begin{figure}[h]
	\centering
	\includegraphics[width=\linewidth]{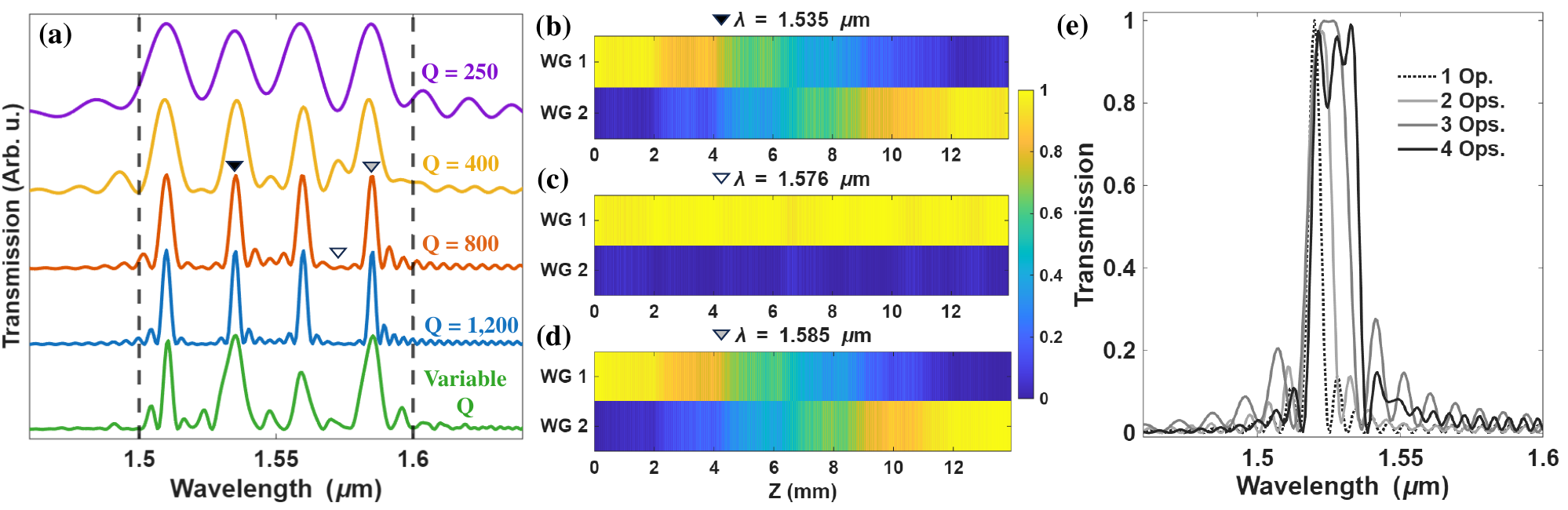}
	\caption{(a) Transmission spectra (cross-waveguide output) of a system with four modulations implemented to work over distinct frequency ranges. The impact of the quality factor on the spectral properties can be seen. Lowest graph (\textit{Variable Q)}: each modulation can be individually varied. Triangles indicate the spectral positions of the propagation maps shown in (b-d). (b-d) Propagation maps for 3 selected wavelength. Light is introduced in the first waveguide. (e) Impact of the spectral superposition of several, up to 4, operations (Ops.).}
	\label{fig2}
\end{figure}

We present first in \fig{fig2}(a) an example where four filters have been implemented to operate at about equi-spaced interval in the $1.51-1.59~{\mu}m$ range. This figure illustrates how the quality factor Q impacts the filters width : Q of about 1,200 results in filters with about 2.5 nm full width at half maximum (FWHM). When the Q factor is reduced, each filter bandwidth is modified accordingly. We illustrate briefly in the lower panel of \fig{fig2}(a) that each filter can have their properties tuned also independently, especially the maximum transmission of each filter can be tuned. Finally we see that for the lowest Q-factor, the filters start merging, resulting in a about sinusoidal transmission pattern, as typically found in (long) integrated couplers. Remarkably, this only occurs over the $1.51-1.59~{\mu}m$ wavelengths range that was encoded; and any other wavelength outside this range would propagate without being affected, as seen in \fig{fig2}(c). Regarding the wavelengths affected by the filter, we see in \fig{fig2}(b,d) that all transitions happen at the same place, meaning that different \textit{non-adiabatic control} operations can be independently collocated; and therefore their arrangement do not require to be sequential. This could be an advantage of classical chirped WBGs that cause an important wavelength dependent delay \cite{Praena2024}, which is absent here. Note that because the design is not further optimized at this point, the interaction in-between filters very close in frequency is not managed. This result in an irregular pattern for the lowest Q-factor.

As different functions interact coherently, avoiding any cross interference requires a careful design and a specific optimization. But on the other hand, this open the possibility to merge together several functions in order to obtain features that differ from the quasi-Lorentzian shape given by the natural bandwidth of a single operation, as described by \eq{eq4}. This is done in \fig{fig2}(e), where one up to seven operations are progressively merged together to result in flat-top square filters of variable bandwidth. Interestingly, the latter can be increased without compromising with the filter slope at its edge. Note however that the combination of several functions can result in increasing ripple at the filter's edge. We chose to present here non-optimized results in order to provide a fair presentation of our framework. Advanced control techniques, including the impact of apodization are briefly discussed in the last section of this paper.

\section{Spatial control}
\begin{figure}[h]
	\centering
	\includegraphics[width=\linewidth]{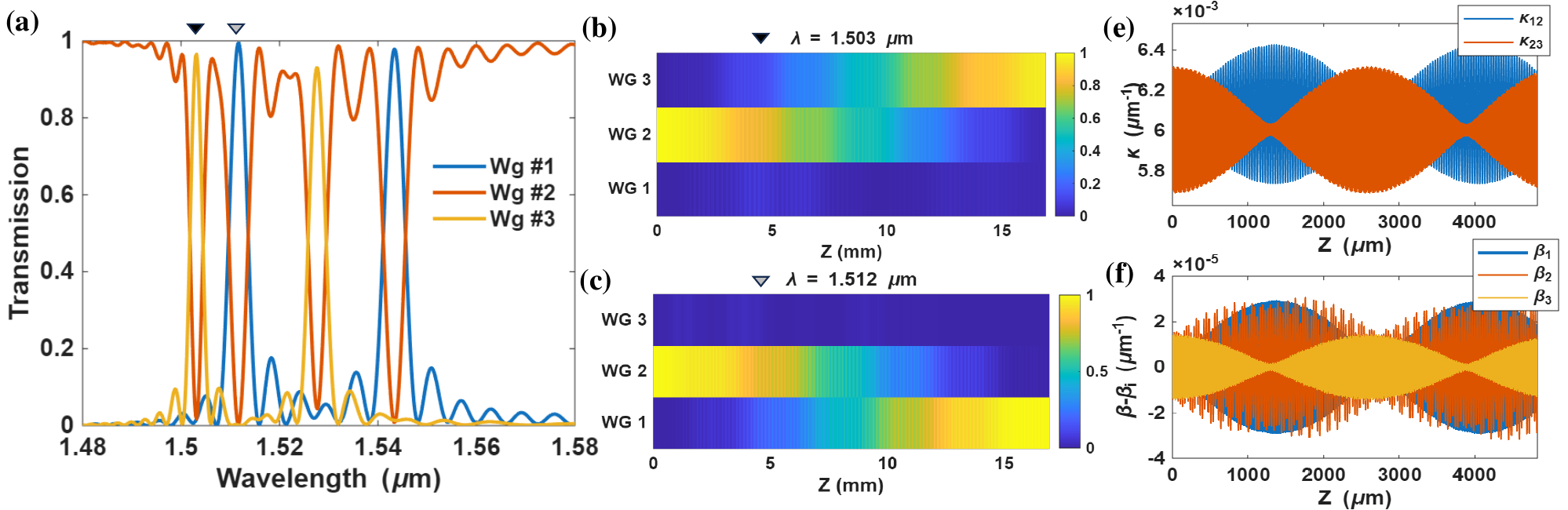}
	\caption{Three-waveguide system demonstrating the routing capability. (a) Spectra depending on the output channel, showing that filters can send light to different outputs. (b–c) Examples of light propagation for two different wavelengths, 1.503 $\mu$m (b) and 1.512 $\mu$m (c). (e-f) Geometrical parameters corresponding to the filter presented in (a). Modulation of the coupling coefficients, directly related to the inter-waveguide spacing, are presented in (e); while modulations of the waveguides' widths is shown in (f).}
	\label{fig3}
\end{figure}

The previous demonstrations were made on a simple two-waveguide system, but the framework of \textit{non-adiabatic control} also works for larger arrays \cite{Sheveleva2022}. In this case, the spectral features we just presented can have also a controlled spatial distribution. We show, as a brief example in \fig{fig3} a three-waveguide system that selectively directs filtered signal to different output channels. This adds spatial features to the filter presented in \fig{fig2}(a). Similarly as for the 2-waveguide system, we see in \fig{fig3}(b-d) that all transitions can be collocated in space.

The corresponding evolution of the parameters, as provided by \eq{eq3}, is shown in \fig{fig3}(e,f). We see that the pattern can become quite complex. Indeed the modulations of the waveguides width (related to the $\beta$ coefficients), and of the inter-waveguide spacing (directly related to the coupling $\kappa_{ij}$ coefficients) differs in both relative phase and amplitude. If the type of modulation that must be applied would be rather straightforward for a 2-waveguide system, the use of \eq{eq3} is essential for larger systems which holds numerous degrees of freedom.

\section{Optimization}

\begin{figure}[h]
	\centering
	\includegraphics[width=\linewidth]{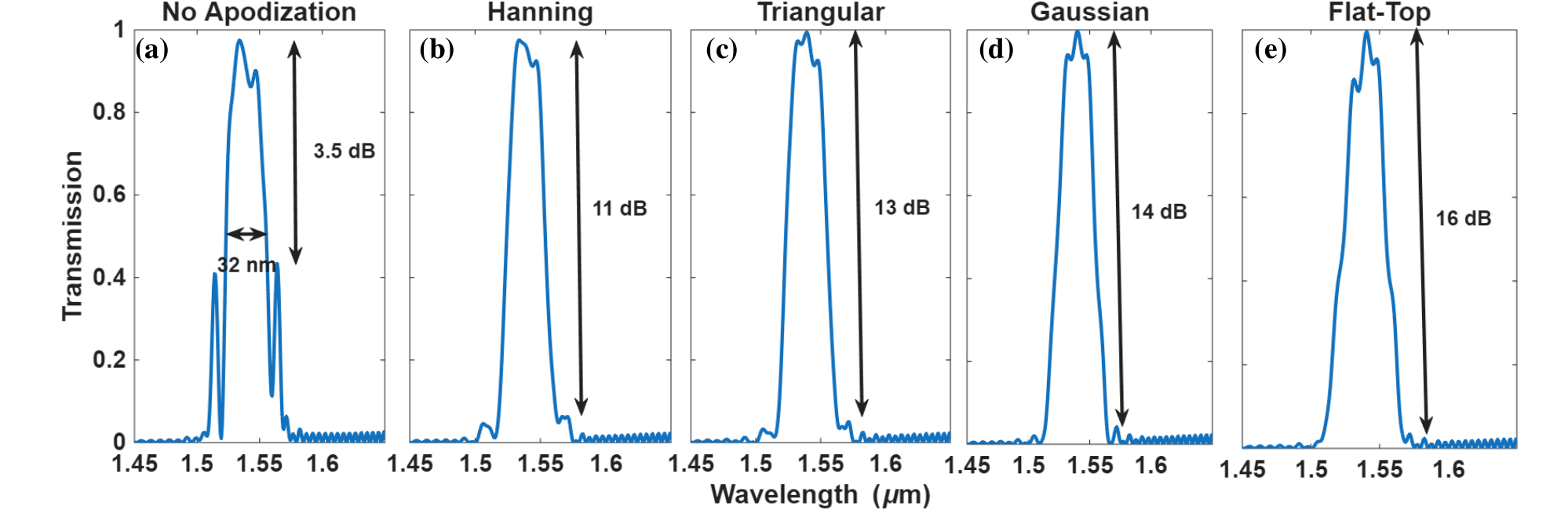}
	\caption{(a) Filter composed of the combination of 9 simple operations, exhibiting an about square shape with 32nm FWHM (the FWHM of a single operation is about 2.5nm). Important ripples are seen at each filter's edge if no apodization is applied. (b-e) Influence of the apodization windows functions on the filter shape. For each situations, the same apodization is applied uniformly to all the operations composing the filter.}
	\label{fig5}
\end{figure}

While the examples we have just presented provide a clear initial understanding, they are not optimized. And therefore they suffers from a few impairments such as, for instance, the characteristic presence of ripples at the filter's edges. Consequently, we would like to briefly discuss how well-known optimization strategies can also be implemented here; and the benefit they could offer. As an illustrative example, we studied in \fig{fig5} the impact of apodization \cite{Cheng2020} over a square filter composed by the interplay of 9 operations.

We see that the ripples can be efficiently suppressed just by applying the same uniform apodization windows to all the operations. As the ripples are suppressed, the overlap between adjacent operations decreases as well, resulting into uneven flat top if their relative spacing is not modified accordingly. Further refinement could be obtained by applying different apodizations on each operation, depending on their spectral position within the filter. In particular, we found that Hanning and Triangular windowings preserve better the center of the filter, while the Flat-Top function must be preferred at the edges. Note that the apodization can be either applied to the final modulation provided by \eq{eq3}, or set directly on the $V^\dagger d_z V$ that serves as input to the latter. In brief, our framework holds many degrees of freedom that are either managed implicitly by the equations we presented, or that can also be explicitly controlled if more advance tuning is required.

\section{Discussion and conclusion}
The  framework that we presented in this article combines seamlessly spectral filtering and spatial routing. Despite the physics at play shares strong similarities with Bragg gratings, the routing capability makes our systems closer to MMI waveguides. However, in contrast to the latter, the design can here be obtained through a deterministic inverse-design procedure (\eq{eq3}).  Note however that, within our framework, higher spatial frequency would mean higher Q factor for the same coupling length. As a result Bragg gratings would always exhibit narrower filters. As an interesting contrast with WBGs, the length scale of the modulations at play during non-adiabatic evolution is very slow. \textit{Non-adiabatic control} and Bragg gratings operate over two completely different length scales; and the two techniques could therefore be combined without suffering from cross-influence. When it come to highly demanding applications, the best strategy would be to combine the \textit{non-adiabatic control} presented here with the Bragg-gratings filters being implemented at the different channel outputs, depending on the requirement. Such an hybrid approach would hold the advantage that the Bragg-filter would then work on a pre-filtered signal where most of the energy is already contained within the correct spectral band. This would hence limit the total amount of energy being filtered out or lost during the filtering. 

\par Regarding the overall filter's architecture, the general configuration is very analogue to add-drop filters. This means that they can be cascaded; and that along with each filters shape which is designed, the complementary operation is also obtained at the same time in another array's outputs. While such operation could be also obtained with a cascade of side coupled cavities, the specific advantage of our proposal is that it allows the shaping of the transfer function of each individual resonance. In particular we can combine several base functions altogether to create large bandwidth filter with steeper slope at the edge, breaking the natural slope-to-bandwidth ratio which is intrinsic to lorentzian resonances; and further optimization such as apodization is possible \cite{Simard2015,Cheng2020}. Interestingly numerous functions can be incorporated without increasing the overall footprint because they can all be collocated in space, allowing more complex and multi wavelength filtering operations without impacting further overall device footprint.

\par Finally, on a more general prespective, couplers are a key building block in integrated photonics. Their power splitting ratio is controlled by the relative phase beating between the super-mode of these 2-waveguides systems, which requires a precise adjustment of the total coupling length. This interferometric requirement hinders the reliable control of more complex functionnality such as 3-by-3 couplers. Indeed, in order to achieve the correct power splitting ratio, the total length of such a coupler could actually become extremely long so as to satisfies all the inteferometric constraint at the same time; and the bandwidth of operation would then shrinks accordingly. Here the modulations of the arrays dictates the power splitting in a very controlled way. It would be also possible to have the same device having different coupling ratios for different frequency bands. We believe multi waveguides systems could offer interesting features to go beyond the classical 2-by-2 couplers, and that the \textit{Non-adiabatic control} could be a interesting asset to go in that direction. One could now reasonably envision to use 3 by 3 -or more complex- couplers as base routing operation for the design of more advanced integrated photonics architecture \cite{Todd2002}.




\bibliography{Biblio_PiCo}






\end{document}